\documentclass[12pt]{article}
\usepackage{epsfig}
\def\etal{{\it et al.}}
\def\@spires#1{\href{http://www-spires.slac.stanford.edu/spires/find/hep/www?j=#1}} 
\catcode`\%=12
\newcommand\adp[3]   {\@spires{ADPHA
		{{\it Adv.\ Phys.\ }{\bf #1} (#2) #3}}
\newcommand\ap[3]    {\@spires{APNYA
		{{\it Ann.\ Phys.\ (NY) }{\bf #1} (#2) #3}}
\newcommand\app[3]   {\@spires{APHYE
		{{\it Astropart.\ Phys.\ }{\bf #1} (#2) #3}}
\newcommand\appol[3] {\@spires{APPOL
		{{\it Acta Phys.\ Polon.\ }{\bf #1} (#2) #3}}
\newcommand\arnps[3] {\@spires{ARNUA
		{{\it Ann.\ Rev.\ Nucl.\ Part.\ Sci.\ }{\bf #1} (#2) #3}}
\newcommand\atmp[3] {\@spires{00203
		{{\it Adv.\ Theor.\ Math.\ Phys.\ }{\bf #1} (#2) #3}}
\newcommand\cpc[3]   {\@spires{CPHCB
		{{\it Comput.\ Phys.\ Commun.\ }{\bf #1} (#2) #3}}
\newcommand\cmp[3]   {\@spires{CMPHA
		{{\it Comm.\ Math.\ Phys.\ }{\bf #1} (#2) #3}}
\newcommand\dmj[3]   {\@spires{DUMJA
		{{\it Duke Math.\ J. }{\bf #1} (#2) #3}}
\newcommand\epjc[3]  {\@spires{EPHJA
		{{\it Eur.\ Phys.\ J. }{\bf C #1} (#2) #3}}
\newcommand\jmp[3]   {\@spires{JMAPA
		{{\it J.\ Math.\ Phys.\ }{\bf #1} (#2) #3}}
\newcommand\jgp[3]   {\@spires{JGPHE
		{{\it J.\ Geom.\ Phys.\ }{\bf #1} (#2) #3}}
\newcommand\jphg[3]   {\@spires{JPAGB
		{{\it J. Phys.\ }{\bf G #1} (#2) #3}}
\newcommand\cqg[3]   {\@spires{CQGRD
		{{\it Class.\ and Quant.\ Grav.\ }{\bf #1} (#2) #3}}
\newcommand\hpa[3]   {\@spires{HPACA
		{{\it Helv.\ Phys.\ Acta }{\bf #1} (#2) #3}}
\newcommand\jhep[3]  
		{{\it J. High Energy Phys.\ }{\bf #1} (#2) #3}
\newcommand\lmp[3]   {\@spires{LMPHD
		{{\it Lett.\ Math.\ Phys.\ }{\bf #1} (#2) #3}}
\newcommand\npa[3]   {\@spires{NUPHA
		{{\it Nucl.\ Phys.\ }{\bf A #1} (#2) #3}}
\newcommand\npb[3]    
		{{\it Nucl.\ Phys.\ }{\bf B #1} (#2) #3}
\newcommand\npps[3]  {\@spires{NUPHZ
		{{\it Nucl.\ Phys.\ }{\bf #1} {\it(Proc.\ Suppl.)} (#2) #3}}
\newcommand\pla[3]   {\@spires{PHLTA
		{{\it Phys.\ Lett.\ }{\bf A #1} (#2) #3}}
\newcommand\plb[3]   
		{{\it Phys.\ Lett.\ }{\bf B #1} (#2) #3}
\newcommand\ppnp[3]  {\@spires{PPNPD
		{{\it Prog.\ Part.\ Nucl.\ Phys.\ }{\bf #1} (#2) #3}}
\newcommand\pr[3]    {\@spires{PHRVA
		{{\it Phys.\ Rev.\ }{\bf #1} (#2) #3}}
\newcommand\pra[3]   {\@spires{PHRVA
		{{\it Phys.\ Rev.\ }{\bf A #1} (#2) #3}}
\newcommand\prb[3]   {\@spires{PHRVA
		{{\it Phys.\ Rev.\ }{\bf B #1} (#2) #3}}
\newcommand\prc[3]   {\@spires{PHRVA
		{{\it Phys.\ Rev.\ }{\bf C #1} (#2) #3}}
\newcommand\prd[3]   
		{{\it Phys.\ Rev.\ }{\bf D #1} (#2) #3}
\newcommand\pre[3]   {\@spires{PHRVA
		{{\it Phys.\ Rev.\ }{\bf E #1} (#2) #3}}
\newcommand\prep[3]  {\@spires{PRPLC
		{{\it Phys.\ Rep.\ }{\bf #1} (#2) #3}}
\newcommand\prl[3]   
		{{\it Phys.\ Rev.\ Lett.\ }{\bf #1} (#2) #3}
\newcommand\ptp[3]   {\@spires{PTPKA
		{{\it Prog.\ Theor.\ Phys.\ }{\bf #1} (#2) #3}}
\newcommand\rmp[3]   {\@spires{RMPHA
		{{\it Rev.\ Mod.\ Phys.\ }{\bf #1} (#2) #3}}
\newcommand\zpc[3]   {\@spires{ZEPYA
		{{\it Z.\ Physik }{\bf C #1} (#2) #3}}
\newcommand\mpla[3]  {\@spires{MPLAE
		{{\it Mod.\ Phys.\ Lett.\ }{\bf A #1} (#2) #3}}
\newcommand\mplb[3]  {\@spires{MPLAE
		{{\it Mod.\ Phys.\ Lett.\ }{\bf B #1} (#2) #3}}
\newcommand\sjnp[3]  {\@spires{SJNCA
		{{\it Sov.\ J.\ Nucl.\ Phys.\ }{\bf #1} (#2) #3}}
\newcommand\jetp[3]  {\@spires{SPHJA
		{{\it Sov.\ Phys.\ JETP\/ }{\bf #1} (#2) #3}}
\newcommand\jetpl[3]  {\@spires{JTPLA
		{{\it Sov.\ Phys.\ JETP Lett.\ }{\bf #1} (#2) #3}}
\newcommand\zetf[3]  {\@spires{ZETFA
		{{\it Zh.\ Eksp.\ Teor.\ Fiz.\ }{\bf #1} (#2) #3}}
\newcommand\yf[3]    {\@spires{YAFIA
		{{\it Yad.\ Fiz.\ }{\bf #1} (#2) #3}}
\newcommand\nc[3]    {\@spires{NUCIA
		{{\it Nuovo Cim.\ }{\bf #1} (#2) #3}}
\newcommand\joth[3]  {\@spires{JOTHE
		{{\it J.\ Operator Theory }{\bf #1} (#2) #3}}
\newcommand\ibid[3]{{\it ibid.\ }{\bf #1} (#2) #3}
\newcommand\ijmpa[3] {\@spires{IMPAE
		{{\it Int.\ J.\ Mod.\ Phys.\ }{\bf A #1} (#2) #3}}
\newcommand\ijmpb[3] {\@spires{IMPAE
		{{\it Int.\ J.\ Mod.\ Phys.\ }{\bf B #1} (#2) #3}}
\catcode`\%=14
\catcode`\|=12

\begin{document}
\begin{titlepage}

\large
\centerline {\bf Conference Summary - Beauty 2000}
\normalsize

\vskip 2.0cm
\centerline {Sheldon Stone\footnote{To be published in Proceedings of Beauty 2000,
Kibbutz Maagan, Israel, September, 2000, edited by S. Erhan,
Y. Rozen, and P. E. Schlein, Nucl. Inst. Meth. A, 2001}}
\centerline {\it Physics Department}
\centerline{\it Email: Stone@phy.syr.edu}
\centerline{\it Syracuse University}
\centerline{\it Syracuse, N. Y. 13244-1130}

\vskip 4.0cm

\centerline {\bf Abstract}
\vskip 1.0cm
I discuss some key issues covered at the Conference including physics goals,
current results on CKM angles, rare decays, status of $|V_{cb}|$ and
$|V_{ub}|$, CKM fits and progress on hardware. 
\vfill
\end{titlepage}

\newpage
\section{Introduction-Physics Goals}

Our goals are to make an exhaustive search for physics beyond the Standard Model
and to precisely measure SM parameters. 
Measurements are necessary on CP violation in $B^o$ and $B_s$ mesons, $B_s$
mixing, rare $b$ decay rates, and mixing CP violation and rare decays in the
charm sector.

The physics we wish to discuss is best described using the Wolfenstein parameterization of the CKM matrix. However, the lowest order approximation is inadequate even now. I advocate using the form accurate to $\lambda^3$ for the real parts and $\lambda^5$ for the imaginary parts. This form is given as:
\begin{equation}
{\begin{array}{ccc} 
1-\lambda^2/2 &   \lambda &  A\lambda^3(\rho-i\eta(1-\lambda^2/2)) \\
-\lambda &   1-\lambda^2/2-i\eta A^2\lambda^4 &  A\lambda^2(1+i\eta\lambda^2) \\
A\lambda^3(1-\rho-i\eta) &   -A\lambda^2&  1  
\end{array}}.
\end{equation}
We know two elements, $\lambda$ is 0.22 and $A$ is $\sim 0.8$. The other two parameters
$\rho$ and $\eta$ are constrained by experimental data. These four parameters
are fundamental constants of nature at the same level of importance as $\alpha$
or $G$.

\begin{figure}[htb]
\centerline{\epsfig{figure=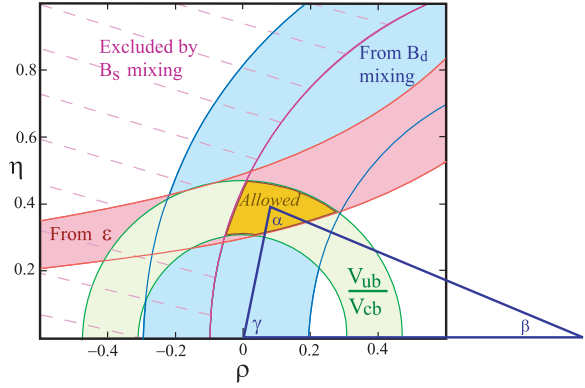,height=2in}}
\vspace{-0.2cm}
\caption{\label{ckm_tri}The CKM triangle shown in the $\rho-\eta$ plane. The
shaded regions show $\pm 1\sigma$ contours given by
$|V_{ub}/V_{cb}|$, neutral $B$ mixing, and CP violation in $K_L^o$ decay ($\epsilon$). The dashed region is excluded by $B_s$ mixing limits.
 The allowed region is defined by the overlap of
the 3 permitted areas, and is where the apex of the CKM triangle sits.}
\vspace{-4mm}
\end{figure}

Constraints from CP violation in $K_L$ decay, $\epsilon$, $|V_{ub}/V_{cb}|$, 
$B_d$ and $B_s$ mixing are shown in Fig.~\ref{ckm_tri}. The bands are shown with $1\sigma$
error bars, with the dominant errors in each case being theoretical. The values
and the errors are controversial. These problems are caused by the basic theory
being applicable to quarks while we make measurements on hadrons
\cite{Stone_HF8}.

To apply QCD to hadrons we have access to two theories and many models. 
The theories are fundamental; they are Heavy Quark Effective Theory,
HQET, and unquenched lattice
gauge theory. HQET is rigorously true for infinite mass quarks. Since we deal with
finite mass quarks, model dependent corrections need to be made. The scheme for
applying these corrections can be straightforward. The lattice calculations
are not complete, due to the effects of not including quark loops (such models
are called ``quenched"). We hold great hopes
for the unquenched calculations.

We would like to make measurements that depend less on models. In many cases
phase measurements are not subject to theoretical uncertainties. For example
lets consider the angles $\alpha$, $\beta$ and $\gamma$ shown in
Fig.~\ref{ckm_tri} \cite{Rosner_rhoeta}.\footnote{These are also called $\phi_1$, $\phi_2$, and
$\phi_3$.} In the Standard Model with three families, these three angles must
sum to 180$^{\circ}$.

\begin{figure}[htb]
\vspace{0.4cm}
\centerline{\epsfig{figure=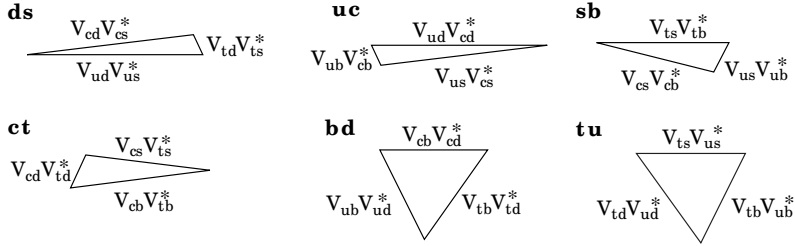,height=1.3in}}
\caption{\label{six_tri}The six CKM triangles. The bold labels, e.g. {\bf ds} 
refer to the rows or columns used in the unitarity relationship.}
\end{figure}

The unitarity of the CKM matrix allows us to construct six relationships.
These may be thought of as the triangles in the complex plane  
shown in Fig.~\ref{six_tri}. (The {\bf bd} triangle is the one depicted in Fig.~\ref{ckm_tri}.)

All six of these triangles can be constructed knowing four and
only four independent angles \cite{silva_wolf}\cite{KAL}\cite{bigis}.
 These 
are defined as:
\begin{eqnarray} \label{eq:chi}
\beta=arg\left(-{V_{tb}V^*_{td}\over V_{cb}V^*_{cd}}\right),&~~~~~&
\gamma=arg\left(-{{V^*_{ub}V_{ud}}\over {V^*_{cb}V_{cd}}}\right), \nonumber\\
\chi=arg\left(-{V^*_{cs}V_{cb}\over V^*_{ts}V_{tb}}\right),&~~~~~&
\chi'=arg\left(-{{V^*_{ud}V_{us}}\over {V^*_{cd}V_{cs}}}\right).\nonumber\\ 
\end{eqnarray}
($\alpha$ can be used instead of $\gamma$ or $\beta$.) Two of the phases $\beta$ and $\gamma$ are probably large while $\chi$ is
estimated to be small $\approx$0.02, but measurable, while $\chi'$ is likely
to be much smaller.

It has been pointed out by Silva and Wolfenstein \cite{silva_wolf} that
measuring only angles may not be sufficient to detect
new physics. For example, suppose there is new physics that arises in 
$B^o-\overline{B}^o$ mixing. Let us assign a phase $\theta$ to this new
physics. If we then measure CP violation in $B^o\to J/\psi K_S$ and eliminate
any Penguin pollution problems in using $B^o\to\pi^+\pi^-$, then we actually
measure $2\beta' =2\beta + \theta$ and $2\alpha' = 2\alpha -\theta$. So while
there is new physics, we miss it, because
$2\beta' + 2\alpha' = 2\alpha +2\beta$ and $\alpha' + \beta' +\gamma
= 180^{\circ}$.

\subsection{A Critical Check Using $\chi$}

The angle $\chi$, defined in equation~\ref{eq:chi}, can be extracted by
measuring the time dependent CP violating asymmetry in the reaction
$B_s\to J/\psi \eta^{(}$$'^{)}$, or if one's detector is incapable of quality
photon detection, the $J/\psi\phi$ final state can be used.  However, in this
case there are
two vector particles in the final state, making this a state of mixed CP,
requiring a time-dependent angular analysis to extract $\chi$, that requires large
statistics.

Measurements of the magnitudes of 
CKM matrix elements all come with theoretical errors. Some of these are hard 
to estimate.
The best measured magnitude is that of $\lambda=|V_{us}/V_{ud}|=0.2205\pm 
0.0018$. 

Silva and 
Wolfenstein \cite{silva_wolf} \cite{KAL}
show that the Standard Model 
can be checked in a profound manner by seeing if:
\begin{equation}
\sin\chi = \left|{V_{us}\over 
V_{ud}}\right|^2{{\sin\beta~\sin\gamma}\over{\sin(\beta+\gamma)}}~~.
\end{equation}
Here the precision of the check will be limited initially by the measurement of
$\sin\chi$, not of $\lambda$. This check can  reveal new physics, even 
if other measurements have not shown any anomalies. There are other checks using
$\left|{V_{ub}\over V_{cb}}\right|$ or $\left|{V_{td}\over 
V_{ts}}\right|$ \cite{Stone_HF8}.

\section{Progress on Required Measurements} 

Table~\ref{table:reqmeas} lists the most important physics quantities and the
decay modes that can be used to measure them. The necessary detector capabilities include
the ability to collect purely hadronic final states, the ability to identify
charged hadrons, the ability to detect photons with good efficiency and
resolution and  excellent time resolution required to analyze rapid $B_s$
oscillations. Measurements of $\cos(2\phi)$ can eliminate 2 of the 4 ambiguities in $\phi$ that are present when $\sin(2\phi)$ is measured. 
\begin{table}[hbt]
\begin{center}
\label{table:reqmeas}
\begin{tabular}{|l|l|c|c|c|c|} \hline\hline
Physics & Decay Mode & Hadron & $K\pi$ & $\gamma$ & Decay \\
Quantity&            & Trigger & sep   & det & time $\sigma$ \\
\hline
$\sin(2\alpha)$ & $B^o\to\rho\pi\to\pi^+\pi^-\pi^o$ & $\surd$ & $\surd$& $\surd$ 
&\\
$\cos(2\alpha)$ & $B^o\to\rho\pi\to\pi^+\pi^-\pi^o$ & $\surd$ & $\surd$& 
$\surd$ &\\
sign$(\sin(2\alpha))$ & $B^o\to\rho\pi$ \& $B^o\to\pi^+\pi^-$ & 
$\surd$ & $\surd$ & $\surd$ & \\
$\sin(\gamma)$ & $B_s\to D_s^{\pm}K^{\mp}$ & $\surd$ & $\surd$ & & $\surd$\\
$\sin(\gamma)$ & $B^-\to \overline{D}^{0}K^{-}$ & $\surd$ & $\surd$ & & \\
$\sin(\gamma)$ & $B^o\to\pi^+\pi^-$ \& $B_s\to K^+K^-$ & $\surd$ & $\surd$& & 
$\surd$ \\
$\sin(2\chi)$ & $B_s\to J/\psi\eta',$ $J/\psi\eta$ & & &$\surd$ &$\surd$\\
$\sin(2\beta)$ & $B^o\to J/\psi K_s$ & & & & \\
$\cos(2\beta)$ &  $B^o\to J/\psi K^o$, $K^o\to \pi\ell\nu$  & &$\surd$ & & \\
$\cos(2\beta)$ &  $B^o\to J/\psi K^{*o}$ \& $B_s\to J/\psi\phi$  & & & 
&\\
$x_s$  & $B_s\to D_s^+\pi^-$ & $\surd$ & & &$\surd$\\
$\Delta\Gamma$ for $B_s$ & $B_s\to  J/\psi\eta'$, $ D_s^+\pi^-$, $K^+K^-$ &
$\surd$ & $\surd$ & $\surd$ & $\surd$ \\
\hline
\end{tabular}
\caption{Required CKM Measurements for $b$'s}
\end{center}
\end{table}

Another interesting way of viewing the physics was given by Peskin
\cite{Peskin}. Non-Standard Model physics would show up as descrepancies among
the values of $(\rho ,\eta)$ derived from independent determinations using CKM
magnitudes ($|V_{ub}/V_{cb}|$ and $|V_{td}/V_{ts}|$), or $B^o_d$ mixing ($\beta$ and
$\alpha$), or $B_s$ mixing ($\chi$ and $\gamma$).

\subsection{Required Measurements Involving $\beta$}

Current measurements of the phase of $B^o-\overline{B^o}$ mixing, expressed
as $\sin(2\beta)$ are listed in Table~\ref{table:s2b}. We note that the world
average value is not yet statistically significant. Placing these constraints
on Fig.~\ref{ckm_tri} would not restrict the ``allowed" region. However, the $b$ factories
have made great starts and the errors should lessen significantly by next
summer. CDF will start in March of 2001 and should vastly improve their measurement.
\begin{table}[hbt]
\centering
\vspace*{2mm}
\begin{tabular}{|l|l|} \hline
 Experiment & $\sin(2\beta)$   \\\hline
BABAR & 0.12$\pm$0.38\\
BELLE & 0.45$\pm$0.45\\
CDF & 0.79$\pm$0.42\\\hline
World Average & 0.42$\pm$0.24\\
\hline
\end{tabular}
\caption{Measured Values of $\sin(2\beta)$\label{table:s2b}}
\end{table}

It is also important to resolve the ambiguities. There are two suggestions on how
this may be accomplished. Kayser \cite{kkayser} shows that time dependent
measurements of the final state
$J/\psi K^o$, where $K^o\to \pi \ell \nu$, give a direct measurement of
$\cos(2\beta)$ and can also be used for CPT tests. Another suggestion is to use
the final state $J/\psi K^{*o}$, $K^{*o}\to K_S\pi^o$, and to compare with
$B_s\to J/\psi\phi$ to extract the sign of the strong interaction phase shift
assuming SU(3) symmetry, and thus determine $\cos(2\beta)$ \cite{isi_beta}.

\subsection{$\alpha$ and $\gamma$}
\begin{figure}[htb]
\vspace{-.3cm}
\centerline{\epsfig{figure=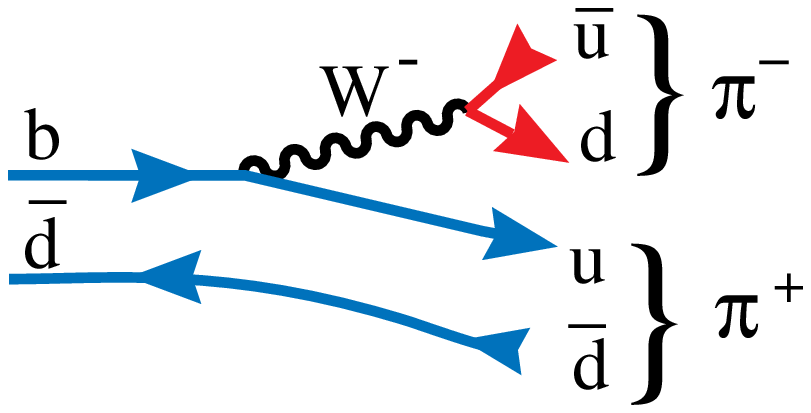,height=1.25in}\epsfig{figure=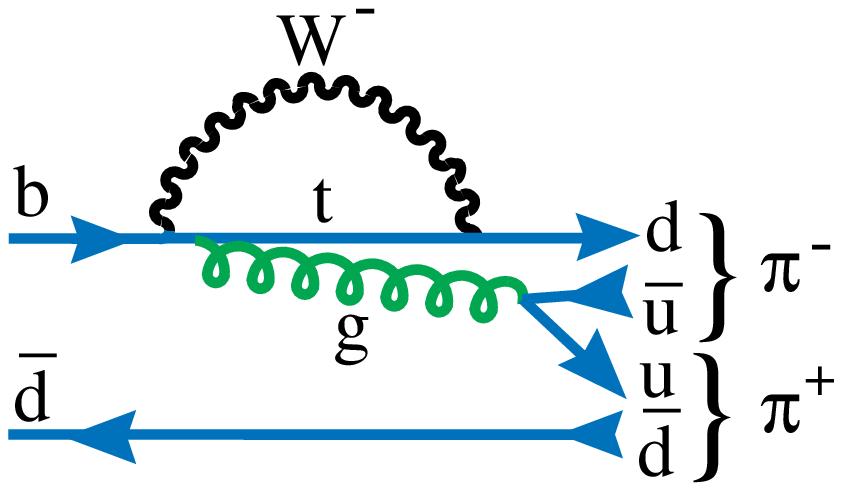,height=1.05in}}
\vspace{-0.2cm}
\caption{\label{pipi} Processes for $B^o\to\pi^+\pi^-$: Tree (left)
and Penguin (right).}
\end{figure}

It is well known that $\sin (2\beta)$ can be measured without
problems caused by Penguin processes using the reaction $B^o\to J/\psi K_S$.
The simplest reaction that can be used to measure $\sin (2\alpha)$ is
$B^o\to \pi^+\pi^-$. This reaction can proceed via both the Tree and Penguin
diagrams shown in Figure~\ref{pipi}.

Current measurements shown in Table~\ref{table:kpi} show a large Penguin component.
\begin{table}[hb]
\centering
\label{table:kpi}
\vspace*{2mm}
\begin{tabular}{|r|c|c|c|c|} \hline
Mode& CLEO & BABAR & BELLE & Average\\\hline
$K^{\pm}\pi^{\mp}$ & $17.2^{+2.5}_{-2.4}\pm 1.2$ & $12.5^{+3.0~+1.3}_{-2.6~-1.7}$ & $17.4^{+5.1}_{-4.6}\pm 3.4$ & $15.5^{+2.0}_{-1.9}$\\
$K^{o}\pi^{\pm}$ & $18.2^{+4.6}_{-4.0}\pm 1.6$ && $16.6^{+9.8~+2.2}_{-7.8~-2.4}$ &  $17.9^{+4.4}_{-3.8}$\\
$K^{\pm}\pi^{o}$ & $11.6^{+3.0~+1.4}_{-2.7~-1.3}$ & & $18.8^{+5.4}_{-4.9}\pm 2.3$  & $13.3^{+2.9}_{-2.6}$\\
$\pi^+\pi^-$ & $4.3^{+1.6}_{-1.4}\pm 0.5$ & $9.3^{+2.6~+1.2}_{-2.3~-1.4}$ & $6.3^{+3.9}_{-3.5}\pm 1.6$ & $5.6^{+1.4}_{-1.2}$\\
\hline
\end{tabular}
\caption{Measured Values of $B$ Branching Ratios into two-body light pseudoscalars.}
\end{table}
The ratio of Penguin {\it amplitude} to Tree {\it amplitude} in the
$\pi^+\pi^-$ channel is about 15\% in magnitude.
Thus the effect of the Penguin must be
determined in order to extract $\alpha$. The only model independent way 
of doing this was suggested by Gronau and London, but requires the measurement
of $B^{\mp}\to\pi^{\mp}\pi^o$ and $B^o\to\pi^o\pi^o$, the latter being rather 
daunting.

There is however, a theoretically clean method to determine $\alpha$.
The interference between Tree and Penguin diagrams can be exploited by
 measuring the time dependent CP violating
 effects in the decays $B^o\to\rho\pi\to\pi^+\pi^-\pi^o$  
as shown by Snyder and Quinn \cite{SQ}.

\begin{figure}[htb]
\vspace{-0.4cm}
\centerline{\epsfig{figure=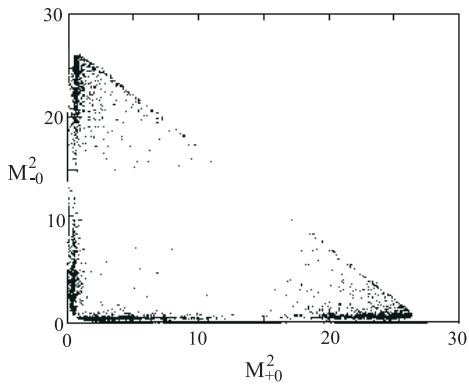,height=2.3in}}
\vspace{-.6cm}
\caption{\label{dalitz} The Dalitz plot for $B^o\to\rho\pi\to\pi^+\pi^-\pi^o$
from Snyder and Quinn.}
\end{figure}

The $\rho\pi$ final state has many advantages. First of all,
it has been seen with a relatively large rate. The 
branching ratio for the $\rho^o\pi^+$ final state as measured by CLEO is 
$(1.5\pm 0.5\pm 0.4)\times 10^{-5}$, and the rate for the neutral
 $B$ final state $\rho^{\pm}\pi^{\mp}$ is  
$(3.5^{+1.1}_{-1.0}\pm 0.5)\times 10^{-5}$, while the $\rho^o\pi^o$ final
state is limited at 90\% confidence level to $<5.1 \times 10^{-6}$
\cite{CLEO_rhopi}. (Bartoldus, at this meeting, reported that BABAR measures
${\cal{B}}\left(B^o\to\rho^{\pm}\pi^{\mp}\right)$ as  
$(4.9\pm 1.3^{+0.6}_{-0.5})\times 10^{-5}$.) These
measurements are consistent with some theoretical expectations \cite{ali_rhopi}.
Furthermore, the associated vector-pseudoscalar
Penguin decay modes have conquerable or smaller branching ratios. Secondly, since the 
$\rho$ is spin-1, the $\pi$ spin-0 and the initial $B$ also spinless, the $\rho$ 
is fully polarized in the (1,0) configuration, so it decays as $cos^2\theta$, 
where $\theta$ is the angle of one of the $\rho$ decay products with the other
$\pi$ 
in the $\rho$ rest frame. This causes the periphery of the Dalitz plot to be 
heavily populated, especially the corners. A sample Dalitz plot is shown in 
Figure~\ref{dalitz}. This kind of distribution is good for maximizing the interferences, which 
helps minimize the error. Furthermore, little information is lost by excluding 
the Dalitz plot interior, a good way to reduce backgrounds.

To estimate the required number of events Snyder and 
Quinn preformed an idealized analysis that showed that a background-free,
flavor-tagged sample of 1000 to
2000 events was sufficient. The 1000 event sample usually yields good results 
for $\alpha$, but sometimes does not resolve the ambiguity. With the 2000 event sample, however, they always succeeded. 

This technique not only finds $\sin(2\alpha)$, it also determines 
 $\cos(2\alpha)$, thereby removing two of the remaining ambiguities. The final
 ambiguity can be removed using the CP asymmetry in $B^o\to\pi^+\pi^-$ and
 a theoretical assumption \cite{gross_quinn}.

\begin{figure}[htb]
\centerline{\epsfig{figure=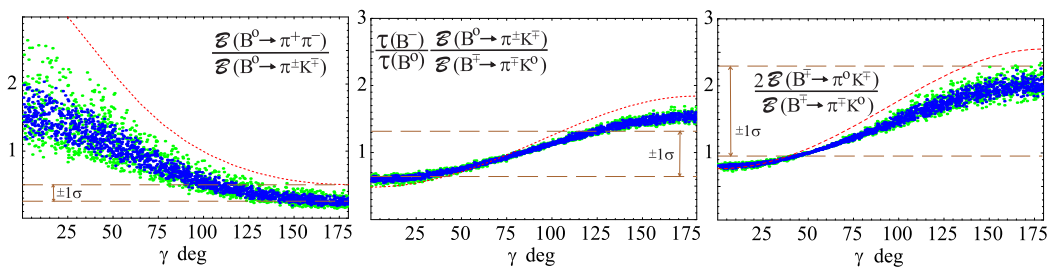,height=1.4in}}
\caption{\label{neub_gamma} Ensembles of model predictions from Beneke \etal ~as
a function of the indicated rate ratios. The dark dots represent ``realistic" models, the lot dots ``conservative" models, the dotted curve the predictions of the leading order result. The $\pm 1\sigma$ bands from current measurements are also shown.}
\end{figure}

Several model dependent methods using the light two-body pseudoscalar decay
rates have been suggested for measuring $\gamma$ The basic idea in all these
methods can be summarized as follows: $B^o\to\pi^+\pi^-$ has the weak decay
phase $\gamma$. In order to reproduce the observed suppression of the decay
rate for $\pi^+\pi^-$ relative to $K^{\pm}\pi^{\mp}$ we require a large
negative interference between the Tree and Penguin amplitudes. This puts
$\gamma$ in the range of 90$^{\circ}$. There is a great deal of theoretical
work required to understand rescattering, form-factors etc... We are left with
several ways of obtaining model dependent limits, due to Fleischer and Mannel \cite{FM},
Neubert and Rosner \cite{NR},
Fleischer and Buras \cite{FB}, and Beneke \etal.~\cite{Beneke}. The latter claim ``model independent"
values for $\gamma$, when precision measurements are made. However, what they
really mean is that by taking into account all models they think are reasonable,
they find a spread of values that can be assigned to a ``theoretical error."
Fig.~\ref{neub_gamma} shows values of $\gamma$
that can be found in their framework, once better data are obtainable.

In fact, it may be easier to measure $\gamma$ than $\alpha$ in a model independent manner. There have been two methods suggested.

(1) Time dependent flavor tagged analysis of $B_s\to D_s^{\pm}K^{\mp}$. This
is a direct model independent measurement \cite{Aleks}.

(2) Measure the rate differences between $B^-\to \overline{D}^o K^-$ and
$B^+\to {D}^o K^+$ in two different $D^o$ decay modes such as $K^-\pi^+$
and $K^+ K^-$. This method makes use of the interference between the tree
and doubly-Cabibbo suppressed decays of the $D^o$, and does not depend
on any theoretical modeling \cite{sad}\cite{gronau}.

\section{Status of CKM Matrix Elements}
How precisely do we know the values of $A$ and the constraints on $\rho$ and
$\eta$ \cite{artuso}?

\begin{figure}[htb]
\centerline{\epsfig{figure=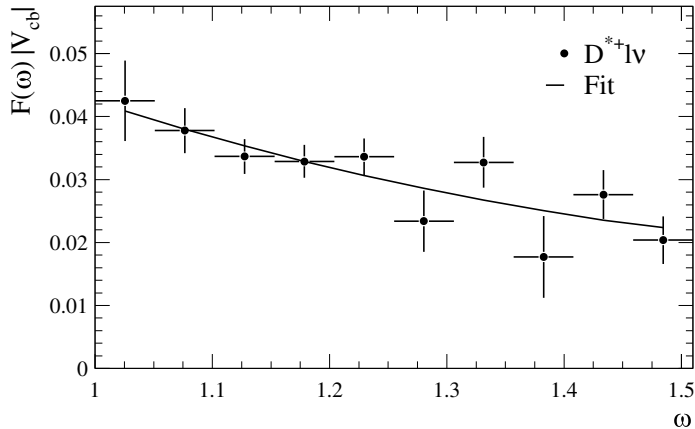,width=3.8in}}
\caption{\label{aleph_vcb} $\overline{B}^o\to D^{*+}\ell^-\bar{\nu}$ from CLEO. 
The data have been
fit to a functional form suggested by Caprini \etal ~The abcissa gives the
value of the product $|F(\omega)\cdot V_{cb}|$.}
\end{figure}

First let us consider $V_{cb}$.
Currently, the most favored
technique is to measure the decay rate of $B\to D^{*}\ell^-\bar{\nu}$ at the
kinematic point where the $D^{*}$ is at rest in the $B$ rest frame. This is 
often referred to as maximum $q^2$ or $\omega =1$, where $\omega=(m^2_{B}+m^2_{D^*}-q^2)/2m_Bm_{D^*}$. Here, according to Heavy
Quark Effective Theory, the theoretical
uncertainties are at a minimum. The CLEO results \cite{cleovcb} are shown in
Fig.~\ref{aleph_vcb}. To extract a value of $V_{cb}$ the data are fit with a function suggested by
Caprini and Neubert \cite{caprini} that is mostly linear with a shape parameter called $\rho^2$, to be determined by the data.

\begin{figure}[htb]
\centerline{\epsfig{figure=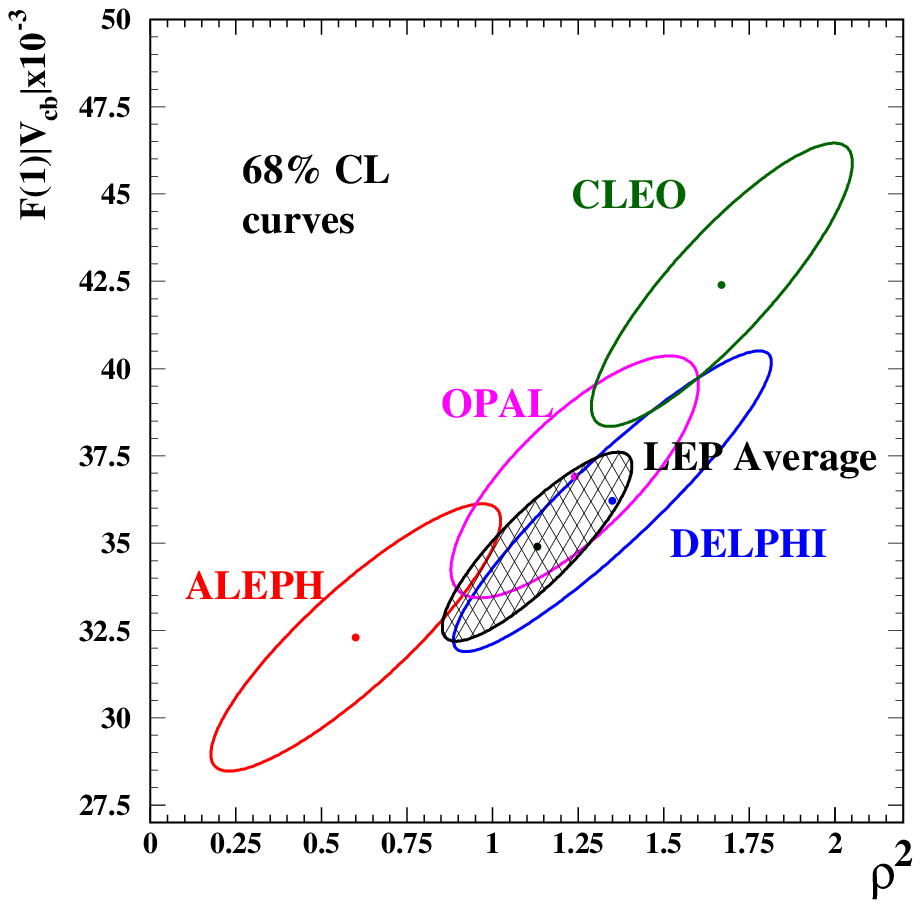,width=3.8in}}
\caption{\label{Vcb_sum} $V_{cb}$ versus $\rho^2$ for LEP and CLEO. The curves are drawn at the 1$\sigma$ confidence level. The cross-hatched region shows the LEP average.} 
\end{figure}

The 68\% confidence level contours for CLEO and the LEP experiments are shown
In Fig.~\ref{Vcb_sum} \cite{Hawkings}. The CLEO data have a slightly higher value for $V_{cb}$ and $\rho^2$ than the LEP data. 

Table~\ref{tab:Vcb} summaries determinations of $F(1)|V_{cb}|$; here,
the first error is the quadrature of the statistical and systematic errors of the experiments, while the third is the common systematic error due to the uncertainties on the $D^o$ and $D^*$ branching ratios.

\begin{table}[th]
\vspace{-2mm}
\begin{center}
\begin{tabular}{|l|c|c|}\hline
Experiment & $F(1)\cdot V_{cb}$ $(\times 10^{-3})$ & $\rho^2$\\
\hline
\hline
LEP& $34.9\pm 0.7 \pm 1.6$ & $1.12\pm 0.08 \pm 0.15$ \\
CLEO& $42.4\pm 1.8 \pm 1.9$ & $1.67\pm 0.11 \pm 0.22$\\
\hline
Average & $37.0\pm 1.3\pm 0.9$ &  $1.30 \pm 0.14$ \\
 \hline
\end{tabular}
\caption{Modern Determinations of $F(1)|V_{cb}|$ using 
$B\to D^*\ell^-\overline{\nu}$ decays
at $\omega = 1$. \label{tab:Vcb}}
\vskip 0.1 in
\end{center}
\end{table}

We now need to obtain a value for $F(1)$ from models. As the quark mass goes to infinity we know that $F(1)$ becomes 1. For finite quark masses we have to calculate corrections as
\begin{equation}
F(1)=1+{\cal O}(\alpha_s/\pi)+\delta_{1/m^2}+\delta_{1/m^3}+...
\end{equation}

Lukes theorem guarantees the absence of the leading $\delta_{1/m}$ terms.
An overall evaluation of the different models gives the form-factor $F(1)=0.913\pm
0.042$ \cite{formfactor}. The value and accuracy have been questioned 
by Bigi \cite{bigivcb}, though his value
and error have changed with time. Eventually we expect to have a good unquenched value from the lattice. Using a value of 0.90$\pm$0.05 I find
\begin{equation}
|V_{cb}|=(40.5\pm 2.5)\times 10^{-3},
\end{equation}
where all the error sources have been added in quadrature. 

The error in $F(1)$ is not a statistical quantity. It is a theoretical error. What does it mean?
Bigi tells us his view \cite{bigivcb}:
 ``In stating a theoretical error, I mean that the real value can lie almost anywhere in this range with basically equal probabilty rather than follow a Gaussian distribution. Furthermore, my message is that I would be quite surprised if the real value would fall outside this range. Maybe one could call that a 90\% confidence level, but I do not see any way to be more
 quantitative." 

Hopefully, in the near future a reliable value with a reliable error will be
given by lattice QCD without using the quenched approximation \cite{simone}.
A good general principle is when dealing with theoretical models it behooves us
to use them to check other measurements before using them to extract important
quantities, such as CKM elements. Furthermore, if they fail the checks they
should not be used. 

There are other ways of determining $V_{cb}$. One method based on QCD sum rules
uses the operator product expansion and the heavy quark expansion
\cite{vub_thy_inc} to
extract a value from the measured semileptonic $b$ branching ratio.
Unfortunately, as stressed by Sachrajda at this meeting, this model fails
to predict the large difference between the measured $\Lambda_b$ and $B^o$
lifetimes. Therefore it does not have proven reliability.

Let us now consider $V_{ub}$.
This is a heavy to light quark transition where HQET cannot be used.
Unfortunately the theoretical models that can be used to 
extract a value from the data do not currently give precise predictions. 

Three techniques have been used. The first measurement of $V_{ub}$ done by CLEO
 and subsequently
confirmed by ARGUS, used only leptons which were more energetic than those that
could come from $b\to c\ell^- \bar{\nu}$ decays \cite{first_vub}. These  
``endpoint leptons'' can occur $b\to c$ background free at the
$\Upsilon (4S)$, because the $B$'s are almost at rest. Unfortunately, there is
only  a small fraction of the $b\to u \ell^-\bar{\nu}$ lepton spectrum that
can be seen this way, leading to model dependent errors. The models used are
either inclusive predictions, sums of exclusive channels, or both
\cite{vubmods}. The average among the models is $|V_{ub}/V_{cb}|=0.079\pm 0.006$,
without a model dependent error \cite{Stone_HF8}.
These models differ by at most 11\%, making it tempting to assign a $\pm$6\%
error. However, there is no quantitative way of estimating the error.

\begin{figure}[bt]
\vspace{-9mm}
\centerline{\epsfig{figure=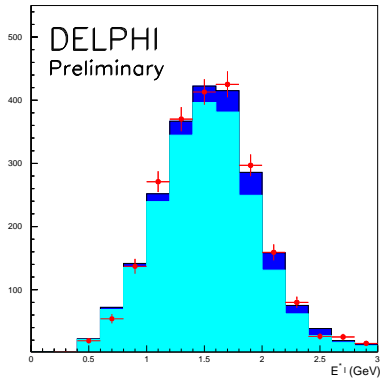,width=2.in}}
\vspace{-4mm}
\caption{\label{delphi_vub}The lepton energy distribution in the $B$ rest
frame from DELPHI. The data have been enriched in $b \to u$ events, and the 
mass of the recoiling hadronic system is required to be below 1.6 GeV. The 
points indicate data, the light shaded region, the fitted background and the 
dark shaded region, the fitted $b \to u \ell \nu$ signal. }
\end{figure}

ALEPH \cite{aleph_vub}, L3 \cite{L3_vub}
and DELPHI \cite{delphi_vub} isolate a class of events where the hadron 
system associated
with the lepton is enriched in $b\to u$ and thus depleted in $b\to c$.     
They define a likelihood that hadron tracks come from $b$ decay by using a large 
number of variables including, vertex information, transverse momentum, not 
being a kaon etc.. Then they require the hadronic mass to be less than 1.6 GeV, which 
greatly reduces $b\to c$, since a completely reconstructed $b\to c$ decay has a 
mass greater than that of the $D$ (1.83 GeV). They then examine the lepton 
energy distribution, shown in Figure~\ref{delphi_vub} for
DELPHI.

The average of all three results as given at this meeting by
Hawkings
\cite{Hawkings}
are $|V_{ub}|=(4.13^{+0.42+0.43+0.24}_{-0.47-0.48-0.25}\pm 0.20)\times
10^{-3}$, resulting in a value for $|V_{ub}/V_{cb}|=0.102 \pm 0.018$, using 
$|V_{cb}|=0.0405\pm 0.0025$. I have two misgivings about this result.
First of all the experiments have to understand the systematic errors
very well. To understand semileptonic $b$ and $c$ decays and thus find
their $b\to u\ell\nu$ efficiency, they employ
different models and Monte Carlo manifestations of these models. To find the error they take half the spread that different models give. This alone may be a serious underestimate. Secondly they use one model,
the OPE model \cite{bigivcb}, to translate their measured rate to a value for $|V_{ub}|$. This model assumes duality,
and there are no successful experimental checks: The model
fails on the $\Lambda_b$ lifetime prediction. Furthermore, the quoted
theoretical error, even in the context of the model, has been estimated by
Neubert to be much larger at 10\% \cite{NeubVub}.


The third method uses exclusive decays.
CLEO has measured the decay 
rates for the exclusive final states $\pi\ell\nu$ and 
$\rho\ell\nu$ \cite{cleo_pirho}. The model of Korner and 
Schuler (KS) was ruled out by the measured ratio of $\rho/\pi$ \cite{vubmods}. 
CLEO has recently
presented an updated analysis for $\rho\ell\nu$ where
they have used several different models to evaluate
their efficiencies and extract $V_{ub}$. These
theoretical approaches include quark models, light cone sum
rules (LCRS), and lattice QCD. The CLEO values are shown
in Table~\ref{tab:Vub}.
\begin{table}[th]
\vspace{-2mm}
\begin{center}

\vskip 0.1 in
\begin{tabular}{|l|c|}\hline
Model & $V_{ub}$ $(\times 10^{-3})$\\
\hline
ISGW2\cite{vubmods} & $3.23\pm 0.14^{+0.22}_{-0.29}$ \\
Beyer/Melnikov\cite{BM} &$3.32\pm 0.15^{+0.21}_{-0.30}$  \\
Wise/Legeti\cite{WL} &$2.92\pm 0.13^{+0.19}_{-0.26}$  \\
LCSR\cite{LCSR} &$3.45\pm 0.15^{+0.22}_{-0.31}$  \\
UKQCD\cite{LCSR} &$3.32\pm 0.14^{+0.21}_{-0.30}$  \\
\hline
\end{tabular}
\end{center}
\caption{Values of $|V_{ub}|$ using 
$B\to \rho\ell^-\overline{\nu}$ and some theoretical models \label{tab:Vub}}
\end{table}

The uncertainties in the quark model calculations (first three in the table)
 are guessed to be
 25-50\% in the rate. The Wise/Ligetti model uses charm data and
SU(3) symmetry to reduce the model dependent errors. The other models estimate their errors
at about 30\% in the rate, leading to a 15\% error in $|V_{ub}|$. 
Note that the models differ by 18\%, but it would be incorrect to assume
that this spread allows us to take a smaller error. At this time it is
prudent to assign a 15\% model dependent error realizing that the errors
in the models cannot be averaged. The fact that the models do not differ
much allows us to comfortably assign a central value 
$|V_{ub}|=(3.25\pm 0.14^{+0.22}_{-0.29}\pm 0.50)\times 10^{-3}$, and
a derived value $|V_{ub}/V_{cb}|=0.080 \pm 0.014$~.
Only the lattice model predictions of UKQCD are used here. More
lattice gauge predictions for the rates in these reactions, at least in some
regions of $q^2$, are promised soon \cite{newlg} \cite{simone} with better errors.
My view is that with experimental checks from measuring form-factors 
and unquenched lattice gauge models the errors will eventually decrease.
 

We can use this latter estimate of $|V_{ub}/V_{cb}|$ along with other
measurements, to get some idea of what the likely values of $\rho$ and 
$\eta$ are.
The $\pm 1\sigma$ contours shown in Figure~\ref{ckm_tri}
 come from measurements of CP violation in $K_L^o$
decay ($\epsilon$), $|V_{ub}/V_{cb}|$ and $B^o$ mixing. Theoretical errors
dominate.  The limit on $B_s$ mixing restricts the range because the ratio of $\Delta m_d/\Delta m_s$ is proportional to $\xi\cdot|V_{ts}/V_{td}|$. The value of $\xi$ is critical for defining the contour. Rosner at this meeting presented a quark model calculation that gives a less restrictive bound than has been used by others \cite{Rosnerxi}.
Some groups have tried to narrow the ``allowed region" by doing maximum
liklihood fits, assigning Gaussian errors to the estimated theoretical
parameters \cite{optimists}. I strongly disagree with this approach. The technique
of Plaszczynski,  \cite{Plas}, while imprecise, is more justifiable.
 
\section{Progress on Hardware}

Modern $b$ decay experiments rely on new technologies. One of the traditionally
more difficult problems has been charged hadron identification. We were saddened this fall by the untimely death of Tom Ypsilantis who along with Jacques Seguinot advanced the important technology of Ring Imaging Cherenkov detectors. Its clear that without Tom and Jacques we would still not know how to separate high momentum kaons and pions, a crucial distinction. We sorely miss Tom.

CLEO III and BABAR have working RICH detectors. The CLEO III detector is a
direct offshoot of developments by Tom and Jacques. It uses LiF radiators and
photon detection with a CH$_4$-TEA gas mixture. BABAR uses quartz bars as
radiators and images the internally reflected light on phototubes. Both systems
are performing well.

\begin{figure}[htb]
\centerline{\epsfig{figure=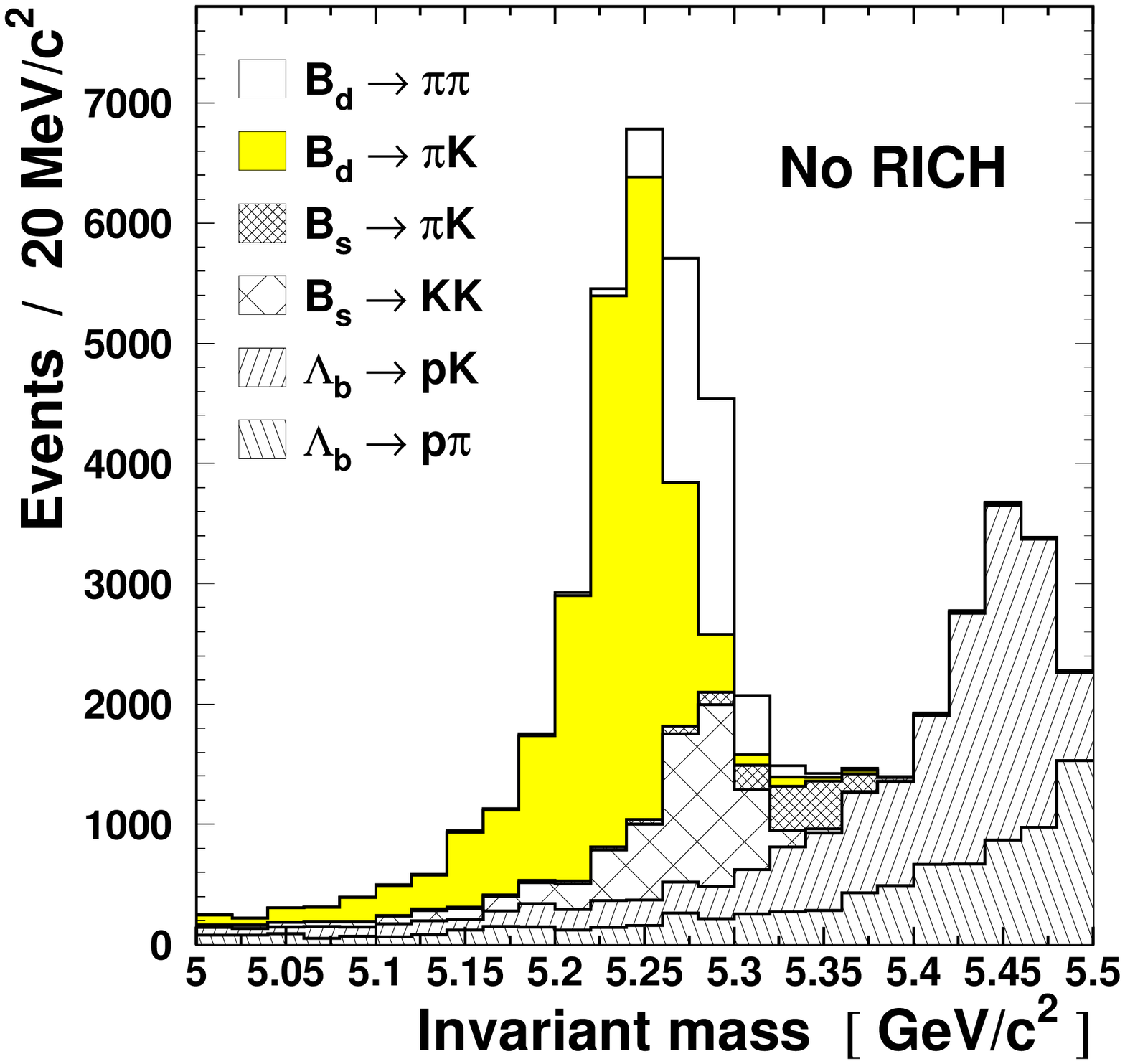,width=2.9in}
\epsfig{figure=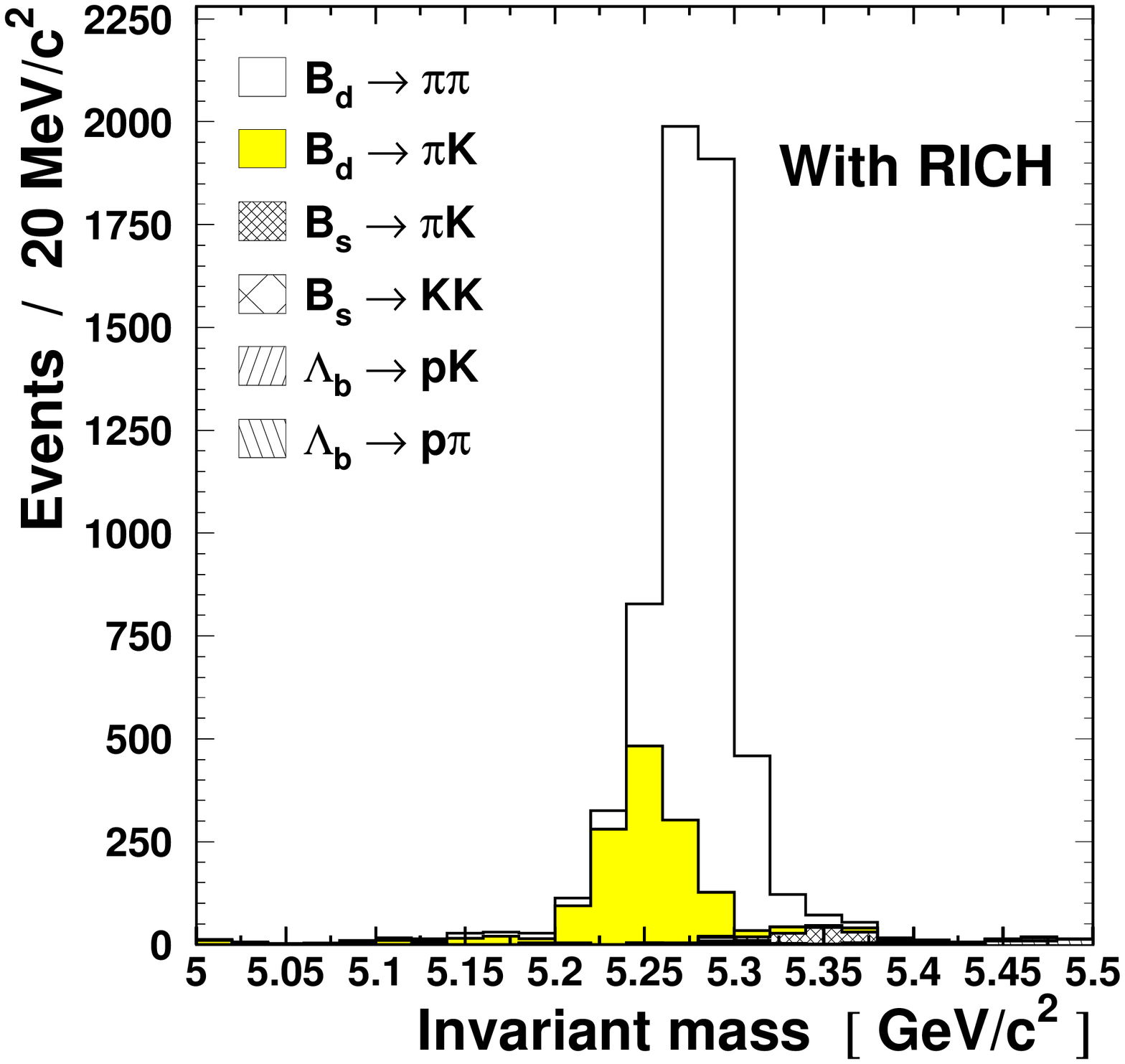,width=2.9in}}
\caption{\label{fig:franz} The reconstructed two-body invariant mass spectrum
in $b$ decay when the two stable particles are interpreted as $\pi^+\pi^-$.
(left) No particle identification information; (right) RICH identification in LHCb.}
\end{figure}

LHCb and BTeV are planning RICH detectors with both gas and aerogel
radiators. The gas radiators are used for high momentum and the aerogel mainly
to distinguish kaons from protons at low momentum. Both groups have decided on
hybrid photodiodes, HPD's, for their baseline. These devices work by converting
a photon on a photocathode, accelerating it with a very high voltage,$\sim$20
kV and crashing it into a silicon pixel detector, that gives a signal of
$\sim$5000 electrons. These devices were chosen over multianode photomultiplier
tubes because they do not require an external focusing system for the
Cherenkov photons and they
currently are about 30\% cheaper. The particle identification power of the RICH
is necessary. Consider, for example, the problem of isolating the
$B^o\to\pi^+\pi^-$ final state from other final states with two stable hadrons.
Without the RICH the signal is swamped by background as shown in
Fig.~\ref{fig:franz}~(left) presented by Franz Muheim on behalf of LHCb. On the
right side the background reduction due to the RICH is evident.

Another example of technological progress is photon detection. Precision photon
detection has been made possible by use of crystal calorimeters. This was
started by Crystal Ball. They had a high performance NaI crystal detector read
out with phototubes, that provided excellent performance. However, they did not
have a magnetic field, so charged particle momenta were not determined. CLEO II
coupled excellent charged particle momentum determination with a CsI crystal
calorimeter, read
out with photodiodes, and having minimal mass in front of the crystals. The
calorimeter has been copied by BABAR and BELLE, with essentially no changes. The
BELLE CsI performs slightly better than CLEO II and the BABAR CSI
slightly worse, in terms of $\pi^o$ mass resolution. 

CMS has developed PbWO$_4$ crystals that can withstand the intense radiation
environment of the LHC. They plan avalanche-photodiode readout. BTeV has
adopted the crystal technology but plans a phototube readout.

Perhaps the most aggressive use of new technology is in triggers and data
acquisition systems for BTeV and the LHC experiments. These are massively
parallel systems that can process a mind boggling amount of data in real time.
John Baines called the ATLAS trigger and DAQ an ``information super-highway,"
and so it is. 

\section{Future Experiments}
Lack of space precludes a more through review here. The $e^+e^-$ experiments,
BABAR and BELLE should see definitive evidence for CP violation in
the $J/\psi K_S$ final state is the next few years. CDF and D0 are now
scheduled to  turn on in 2001. CDF already has shown that it has the
potential to measure
CP in $J/\psi K_S$ \cite{borto}, and
promises to measure $B_s$ mixing. 

To over constrain the CKM matrix and look for new physics all the quantities
listed in Table~\ref{table:reqmeas} requires, however, much larger samples
of $b$-flavored hadrons, and detectors capable of tolerating large interaction
rates and having excellent lifetime resolution, particle identification
and $\gamma/\pi^o$ detection capabilities. The large $b$ rates, including the
$B_s$, are available only at hadron colliders. Two dedicated experiments are 
now approved LHCb and BTeV. These experiments have the potential to measure
all of these parameters and others including rare $b$ decays and CP violation,
rare decays and mixing in charm. Much interesting information will come from ATLAS and CMS especially in dilepton decays.

\section{Conclusions}
This year has brought many interesting physics results with the amazing turn on
of the asymmetric $b$ factories and many more will be forthcoming in the near
future. The panel discussion led by Fred Gilman on the last day concerning 
current values of CKM matrix elements and parameters was most illuminating.
BTeV has joined LHCb as an approved experiment and there will be good
competition in the ultra sensitive $b$ measurements in the LHC era.

\section{Acknowledgements}
I thank Peter Schlein and Yoram Rozen for organizing a most interesting
conference. The best quote was by Yoram who entered the Jordan river on a kyak
with Peter. We next saw Yoram floating down the rapids without Peter screaming
``throw me a paddle." Well starting out on $b$ physics he wasn't the
first one who ended up the creek without a paddle, but someone did toss him a
paddle and he moved out of danger.

I am grateful to all of those who gave presentations, with special thanks
to Marina Artuso, Franz Muheim, Jon Rosner and Chris Sachrajda.

This work was
supported by the National Science Foundation.

\end{document}